\documentclass[aip,amsmath,amssymb,reprint]{revtex4-1}
\usepackage{graphicx,epsf,amsmath,amssymb,wasysym,verbatim,color,cleveref}

\begin{document}
\title[Interplay between degree and Boolean rules in Boolean networks]
{Interplay between degree and Boolean rules in the stability of Boolean networks}
\author{Byungjoon Min}
\email{min.byungjoon@gmail.com}
\affiliation{Department of Physics, Chungbuk National University, Cheongju, Chungbuk 28644, Korea}
\affiliation{Research Institute for Nanoscale Science and Technology, Chungbuk National University, 
	Cheongju, Chungbuk 28644, Korea}
\date{\today}

\begin{abstract}
Empirical evidence has revealed that biological regulatory systems are controlled by 
high-level coordination between topology and Boolean rules. In this study, we study
the joint effects of degree and Boolean functions on the stability 
of Boolean networks. To elucidate these effects, we focus on i) the correlation 
between the sensitivity of Boolean variables and the degree, and ii) the coupling 
between canalizing inputs and degree. We find that negatively correlated 
sensitivity with respect to local degree enhances the stability of Boolean networks 
against external perturbations. We also demonstrate that the effects of canalizing 
inputs can be amplified when they coordinate with high in-degree nodes. Numerical 
simulations confirm the accuracy of our analytical predictions at both the node 
and network levels.
\end{abstract}
\maketitle

\begin{quotation}
How to predict and assess the stability of the Boolean networks is an important 
problem of much interest. Recently, an exhaustive search of real-world 
biological circuits has revealed the high-level coordination between 
network topology and Boolean variables. However, the role of the correlations 
between structural and dynamic properties is still not fully understood.
Here, we study the stability of random Boolean networks with intertwined 
the degree and Boolean functions. We demonstrate that analysis based on the naive 
mean-field approach may fail to predict dynamical consequences in Boolean networks
with intertwined structural and functional properties, which are often observed 
in real-world biological systems. 
\end{quotation}

\section{Introduction}

The random Boolean network proposed by Kauffman in $1969$~\cite{kauffman1969} 
has been widely used in physics, biology, and computer science for modeling biological 
regulatory systems in an abstract manner~\cite{karlebach2008,albert2008,li2004,
thomas2001,shmulevich2002}. Many functions in living systems can be
modeled by Boolean networks, including genetic regulation~\cite{kauffman1969,
li2004,thomas2001}, neural firing~\cite{rosin2013}, and social activity~\cite{sayama2013}. 
The dynamical patterns of Boolean networks fall into two phases, namely 
stable and unstable (chaotic) phases. In a stable phase, most nodes rapidly 
reach a steady state and remain unchanged. In contrast, in an unstable phase most 
nodes change their states in a chaotic manner.
It has been suggested that many biological regulatory systems ranging from genetic 
systems to neural systems tend to hover near the borderline between these two phases,
achieving both stability and evolvability~\cite{kauffmanbook,munoz2018,krotov2014}. 
Empirical evidence supports the hypothesis that biological networks remain near 
criticality~\cite{kauffmanbook,daniels2018}, especially for knockout 
experiments for single genes in {\it Saccharomyces cerevisiae}~\cite{serra2004} 
and gene expression dynamics in macrophages~\cite{nykter2008}.

Theoretical predictions regarding the dynamics of Boolean networks are 
based on stability against perturbations~\cite{derrida1986,luque1997}. 
While damage caused by perturbations dies out quickly in a stable phase, it 
can spread through an entire system in an unstable phase. 
Pioneering research on the theoretical prediction of the stability
of random Boolean networks has revealed that the mean degree $\langle k \rangle$ 
of a network and the mean bias $\langle p \rangle$ of Boolean functions typically 
determine the location of criticality~\cite{derrida1986,luque1997}. 
Since then, many studies have attempted to assess the effects of structural 
and dynamical features, including scale-free structures~\cite{aldana2002,lee2008}, 
noise~\cite{peixoto2009,villegas2016}, multi-level interactions~\cite{cozzo2012}, 
asynchronous updates~\cite{klemm2005}, continuous dynamics~\cite{ghanbarnejad2011},
veto functions~\cite{ebadi2014}, bipartite interactions~\cite{lee2012}, 
knockout~\cite{boldhaus2010,wang2018}, adaptive dynamics~\cite{goudarzi2012}, and 
canalizing functions~\cite{moreira2005}.

Recently, an exhaustive search of real-world networks revealed that biological 
regulatory circuits achieve stable and adaptive functionality based on the 
interplay between logical variables and causal structures \cite{daniels2018}. 
To be specific, anti-correlated sensitivity to the local degree of nodes 
and abundant canalizing functions were observed in the real-world biological 
networks across diverse regulatory systems \cite{daniels2018,moreira2005}.
Therefore, it is expected that real-world biological networks 
achieve near-critical behavior in a distinguished way than random Boolean networks 
without any correlation between Boolean rules and topology.
There have been several attempts to study the role of the correlations 
between structural and dynamic properties in Boolean 
networks \cite{shmulevich2004,squires2014,pomerance2009},
but this role is still not fully understood, particularly for the
correlations between node degree and Boolean functions.

In this study, we analyze the stability of a random Boolean network
incorporating interplay between network topology and Boolean variables.
Specifically, we aim to assess the effects of the correlation between 
local node degree and Boolean functions in terms of
sensitivity~\cite{shmulevich2004,squires2014} and canalizing 
inputs~\cite{moreira2005}. In this paper, we elucidate the role of the 
coupling between local topology and Boolean rules in promoting the stability of 
Boolean networks. Specifically, we demonstrate that negatively 
correlated sensitivity to degree enhances the stability of Boolean 
networks. Additionally we find that coordination between high-degree 
nodes and canalizing inputs can enhance stability. Numerical simulations 
are conducted to verify our analytical predictions, revealing excellent agreement.

\section{Theory}

\subsection{Semi-annealed approximation}

A Boolean network consists of a set of nodes whose states are binary 
(i.e., {\it on} or {\it off}). Directed links between nodes in a Boolean network 
represent the regulatory interactions. We denote in- and out-degrees of
node $i$ as $k_i^{in}$ and $k_{i}^{out}$. We assign the bias $p_i$ for 
node $i$ which indicates the probability of a Boolean function to
return {\it on} state. A Boolean function or truth table for every 
combination (total of $2^{k_{i}^{in}}$ combinations) of inputs 
is assigned according to the bias $p_i$. Here we consider a single fixed 
network and randomly reassigning Boolean inputs to all nodes at each time step, 
so called semi-annealed approximation.

Starting from an initial state selected at random, the state of each node is 
updated synchronously according to its Boolean function and input signals. 
After a transient period, the dynamics of the Boolean network eventually 
arrives at a set of restricted patterns among of total of $2^N$ possible 
states, where $N$ is the number of nodes. To simulate a perturbation, a small 
fraction of the nodes are randomly selected and flipped. To check the network 
responses to such perturbations, we define the stability of the network as 
its ability to eliminate damage. In a stable phase, nodes flipped by a 
perturbation quickly return to their initial states. However, in an unstable 
phase, the majority of the nodes in a system fall under the influence of 
perturbations and evolve to exhibit chaotic dynamics.

To quantify the stability of a Boolean network, we measure the (normalized) 
Hamming distance $H$ between the initial and final states following a perturbation.
We define the state of the nodes as $\vec{s}=\{s_1,s_2,\cdots,s_N\}$, where 
$s_i \in \{0,1\}$. The average Hamming distance between an initial ($t_o$) and 
final ($t$) state is defined as 
\begin{align}
H = \frac{1}{N} \sum_i | s_i(t) - s_i(t_o) |.
\end{align}
While $H$ remains at zero in a stable phase within the thermodynamic limit as 
$N \rightarrow \infty$, it takes on non-zero values in unstable phases. 
Therefore, $H$ represents the degree of network instability.

For a given bias $p_i$ of node $i$, the probability that node $i$ changes its 
state when one of its input changes, which is referred to as sensitivity, is 
defined as $q_i=2 p_i(1-p_i)$. We define $H_{i}$ as the probability that the 
state $s_i$ of node $i$ changes based on a change in one of its neighbors. 
For a locally tree-like network, we can derive the following self-consistency 
equations for a set of Hamming distances $H_{i}$ for each 
node $i$~\cite{pomerance2009,squires2012}:
\begin{align}
H_i =  q_i \left[ 1- \prod_{j \in \partial i} (1-H_{j}) \right],
\label{eq:hi}
\end{align}
where $\partial i$ represents the set of neighbors of node $i$.
When iterating Eq.~\ref{eq:hi} from an initial value of $H_{i}$,
it converges to a fixed point.
We can then obtain the average Hamming distance for an entire network 
as follows: 
\begin{align}	
\langle H \rangle = \frac{1}{N}\sum_i H_i.
\label{eq:hih}
\end{align}
Note that Eq.~\ref{eq:hi} can be interpreted as a percolation process with 
an occupation probability of $q_i$ \cite{squires2012,bmin,bmin2}.

We linearize Eq.~\ref{eq:hi} around the null vector for $H_i$,
calling $\epsilon_i$ the value of $H_i$ in this linearization.
By expanding $H_i$ near a small value of $\epsilon_i$, we get 
\begin{align}
\epsilon_{i} 
=  q_i   \sum_{j } \mathcal{A}_{ij} \epsilon_{j},  
\label{eq:epsilon}
\end{align}
where $\mathcal{A}_{ij}$ are the elements of the adjacency matrix.
It should be noted that we neglect second- and higher-order terms.
Next, the critical point can be identified by calculating the inverse of 
the principal eigenvalue $\Lambda$ of the matrix $\mathcal{Q}$ as follows:
\begin{align}
\mathcal{Q}_{ij} = q_i \mathcal{A}_{ij}.
\end{align}
By using Eqs.~2-5, we can calculate the stability and critical point
for a fixed network structure.

\subsection{Annealed approximation}

We next introduce an annealed approximation for randomly
reassigning Boolean inputs and network links at each time step,
neglecting the fact that the Boolean functions and the network 
structure are quenched. Then we can treat the $H_{i}$ value for 
each node as the same value $H_{a}$. In this approximation, analysis 
at a single node level is no longer possible, but we can easily 
compute the stability of a Boolean network by solving a single 
equation for $H_{a}$ with given degree and sensitivity distributions. 
We obtain the following equation for a degree distribution $P(k^{in},k^{out})$, 
where $k^{in}$ and $k^{out}$ are the in and out degree, respectively: 
\begin{align}
H_{a}= 1- \sum_{k^{in},k^{out}} \frac{k^{out} P(k^{in},k^{out})}{\langle k^{out} \rangle} q(k^{in}) (1-H_{a})^{k^{in}},
\label{eq:ann}
\end{align}
where $q(k^{in})$ is the sensitivity distribution as a function of in-degree.
Assuming that $k^{in}$ and $k^{out}$ are uncorrelated, we get 
\begin{align}
H_{a}= 1- \sum_{k^{in}} P(k^{in}) q(k^{in}) (1-H_{a})^{k^{in}}. 
\label{eq:ann_in}
\end{align}
We now define $f(H_{a})=1- \sum_{k^{in}} P(k^{in}) q(k^{in}) (1-H_{a})^{k^{in}} - H_{a}$.
By applying the linear stability criterion, 
the critical point can be identified by the condition $f'(0)=0$, 
which yields
\begin{align}
\sum_{k^{in}} k^{in} P(k^{in}) q(k^{in}) =1. 
\label{eq:pc}
\end{align}
Assuming that the sensitivity has no correlation with the 
degree, we can recover the well-known prediction of the critical 
point $\langle k^{in} \rangle = 1/\langle 2p(1-p) \rangle$ \cite{derrida1986,luque1997}.

\section{Results}

We analyze the effects of the correlation between node degree and sensitivity using 
the general framework described above. First, we constructed an Erd\"os-R\'enyi (ER) 
graph and assign the sensitivity $q_i=2p_i(1-p_i)$, where $p_i$ is the bias. 
Here we used a directed version of ER graphs meaning that we assign 
directionality randomly at each link. In addition, we assume that the degree distribution 
$P(k^{in},k^{out})$ of ER graphs follows the Poisson distribution as the 
expected distribution for $N \rightarrow \infty$.
We consider three representative cases of the coupling 
between sensitivity and node in-degree: uncorrelated (UC), positively correlated (PC), and 
negatively correlated (NC). For UC case, we assign the same $\langle p \rangle$ to each 
node to obtain uncorrelated coupling. For the PC case, we assign the linearly correlated bias 
$p_i$ of node $i$ to its in-degree $k_i^{in}$ as $p_i = C_{P} k_i^{in}$. Here, $C_{P}$ determines 
the average bias for a given mean in-degree as $\langle p\rangle=C_{P} \langle k^{in} \rangle$. 
In contrast, for the NC case, $p_i$ is assigned as $p_i = - C_{N} k_i^{in} + 1/2$, where 
$C_{N}$ determines the average bias as $\langle p\rangle=-C_{N} \langle k^{in} \rangle + 1/2$. 
The factor of $1/2$ ensures that the maximum value of bias is $1/2$. 
The exact linear relationships in the PC and NC cases do not sustain all possible ranges of 
$\langle p\rangle$ because $0\le p_i \le 1/2$. However, the range of linear dependency 
is still sufficiently broad to examine the impact of the correlation. 
By substituting all of these parameters into Eq.~\ref{eq:ann_in},
we can derive the self-consistency equations for $H_{a}$ for the three coupling cases as follows:
\begin{align}
\label{eq:er}
\text{UC:} &\ H_{a} = 2p(1-p)(1-e^{-\langle k^{in} \rangle H_{a}}), \\
\text{PC:} &\ H_{a} = 2C_{P} \langle k^{in} \rangle \Big\{ 1-C_{P}(1+\langle k^{in} \rangle) \nonumber \\
& + (1-H_a) e^{-\langle k^{in} \rangle H_{a}} \left[C_p +C_p \langle k^{in} \rangle (1-H_a) -1 \right] \Big\}, \nonumber \\
\text{NC:} &\ H_{a} = \frac{1}{2} -2 C_N^2 \langle k^{in} \rangle (1+\langle k^{in} \rangle) -e^{- \langle k^{in} \rangle H_a}  \nonumber \\
&\times \left[\frac{1}{2} -2C_N^2 (1-H_{a}) \langle k^{in} \rangle(1+ \langle k^{in} \rangle - \langle k^{in} \rangle H_{a}) \right]. \nonumber
\end{align}
By solving these self-consistency equations, we can obtain the average 
Hamming distances and identify the critical points.

\begin{figure}
\includegraphics[width=\linewidth]{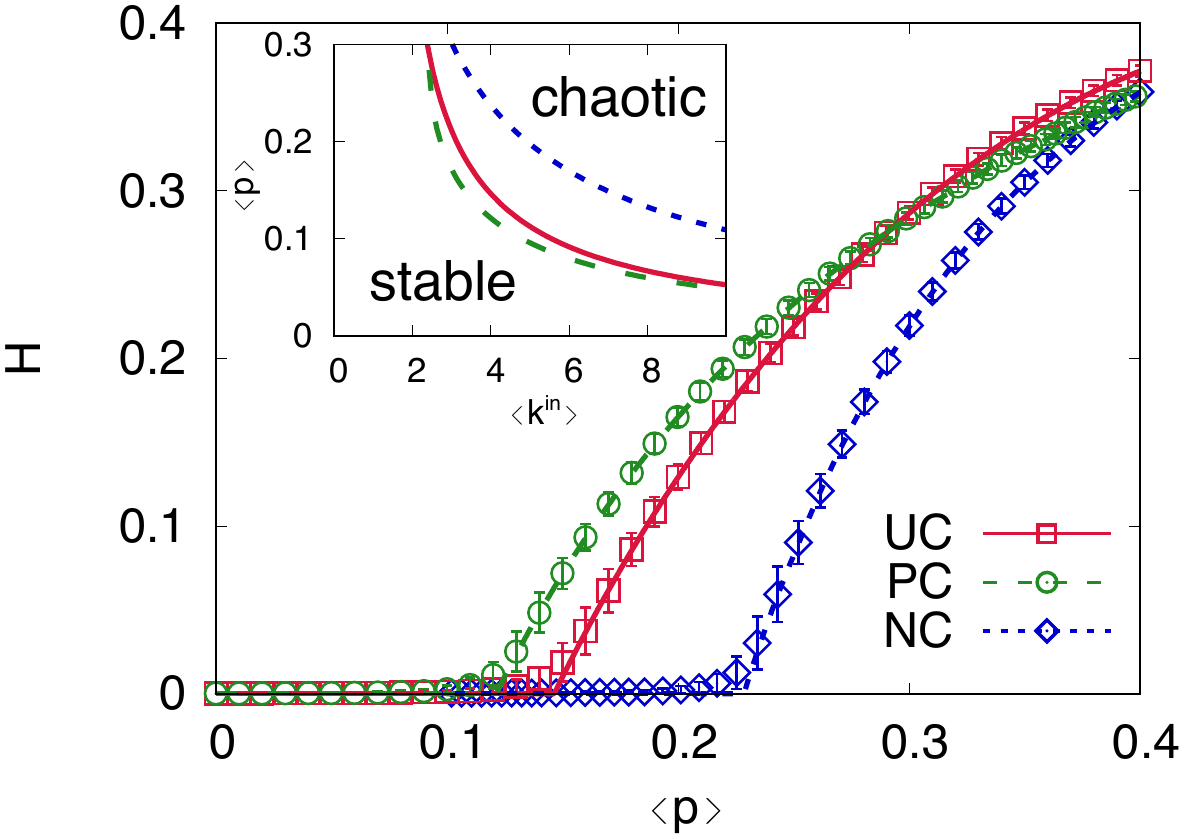}
\caption{
Average Hamming distances of random Boolean networks with
three types of correlated coupling (UC, PC, and NC). 
Analytical predictions (lines) and numerical results (symbols) are presented 
together. We use ER networks with $\langle k^{in} \rangle=\langle k^{out} \rangle=4$ 
and $N=10^4$, and generate $10^3$ different realizations. (inset) Phase 
diagram of the stable and chaotic phases for the three correlated 
Boolean models.
}
\label{fig:er}
\end{figure}

We implemented numerical simulations on ER networks with 
$\langle k^{in} \rangle = \langle k^{out} \rangle=4$ 
without any degree-degree correlation. We assigned the biases and corresponding
Boolean variables according to the process described above. From initial states 
selected randomly, the state of each node are updated synchronously according to 
the Boolean variables. After a transient period, the trajectory will fall into a 
point or cycle attractor since the dynamics are deterministic. 
To simulate a perturbation, we flipped a fraction of $0.01$ of the nodes by force. 
When the system reached a point or cycle attractor again following the 
perturbation, we measured the Hamming distance over all nodes.

In Fig.~\ref{fig:er}, we compare the analytical predictions from Eq.~\ref{eq:er}
to the numerical simulations. The agreement between the theory and the simulations
is excellent. We find that a negatively correlated sensitivity to degree 
enhances stability when comparing the UC and PC cases. For the NC case, the transition point of 
mean bias $\langle p\rangle_c$ is delayed and $H$ is lower than in the other cases.
In contrast, the PC case exhibits an enlarged chaotic region compared to the other cases, 
making it more vulnerable to perturbations. These results demonstrate that 
the correlation between sensitivity and degree can significantly affect
the stability in terms of the location of the critical 
point and the size of Hamming distance.

In order to examine the effect of the correlation in more realistic 
circuits, we consider two examples beyond ER networks: i) an empirical regulatory 
network \cite{conroy} and ii) a network with a heavy-tailed out-degree 
distribution. First, we consider a real-world regulatory network of CD4$^+$ T-cell
with $N=188$ and $\langle k \rangle=3.73$ \cite{conroy}. Each node and link in the 
network respectively represents proteins and protein-protein interactions. 
We assume every link in the network is bi-directional. 
We choose the CD4$^+$ T-cell network as a representative example because 
the network is the second largest example in the all tested networks 
in the survey of empirical Boolean networks \cite{daniels2018} and also has a sufficient 
link density to examine the stability of the network.
Second, we study networks with scale-free out-degrees and Poisson 
distributed in-degrees. The rationale behind these networks is that real transcriptional 
regulatory networks commonly possess a heavy-tailed out-degree distribution and 
a short-tailed in-degree distribution \cite{guelzim,dobrin}. We constructed $10^3$ different
network realizations with size $N=10^4$ having scale-free out-degrees with the degree 
exponent $\gamma=2.5$ and Poisson distributed in-degrees 
with $\langle k_{in} \rangle =4.72\ldots$, according to the configuration model.
We implemented numerical simulations on the two networks 
with three couplings as described above.

\begin{figure}
\includegraphics[width=\linewidth]{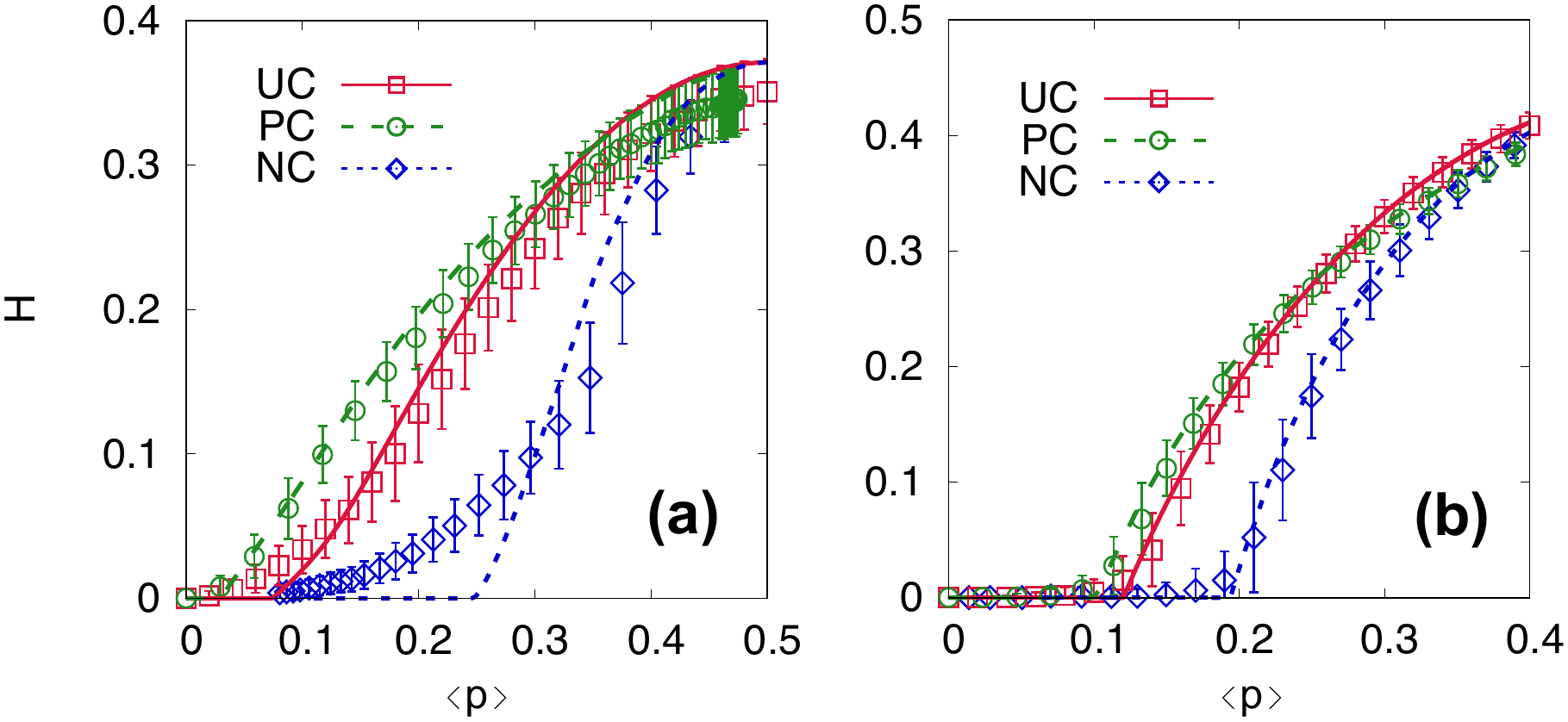}
\caption{
(a) Average Hamming distances $H$ of a real-world Boolean network
for a CD4$^+$ T-cell with $N=188$ and $\langle k \rangle =3.73$.
(b) Average Hamming distances $H$ of Boolean networks 
with a scale-free out-degree distribution with the degree exponent $\gamma=2.5$
and a Poisson in-degree distribution with $\langle k_{in} \rangle \approx 4.72$.
We generate $10^3$ different network realizations with $N=10^4$.
We consider three types of correlated coupling (UC, PC, and NC).
Analytical predictions (lines) and numerical results (symbols) are presented together. 
}
\label{fig:real}
\end{figure}

In Fig.~\ref{fig:real}, we show the analytical predictions and the numerical 
simulations (symbols) for the CD4$^+$ T-cell network and scale-free out-degree distribution
network. The theory (lines) based on Eqs.~\ref{eq:hi} and \ref{eq:hih} shows well agreement 
with the simulations in a qualitative manner. Deviation from the theory in the 
CD4$^+$ network shown in Fig.~\ref{fig:real}(a) is due to the short loops and 
small size of the real network. 
We find again a negatively correlated sensitivity to degree 
increases the stability of Boolean dynamics when comparing the UC and PC cases. 
We confirm the delayed transition $\langle p \rangle_c$ to chaotic phase and 
lower $H$ for the NC coupling than the other cases. On the other hand, the 
PC coupling shows less stable behaviors to perturbations. There is an inversion
between $H$ for UC and PC cases for $\langle p \rangle \gtrsim  0.25$, far 
above $\langle p \rangle_c$ in Fig.~\ref{fig:real}(b). In this region, low degree nodes
with low bias in PC coupling suppress $H$, so that $H$ for PC is above 
that for UC. The similar crossing behaviors are often observed when one considers degree
correlations such as assortative mixing \cite{assortative} and interlayer degree 
correlations \cite{interlayer}.

\begin{figure}
\includegraphics[width=\linewidth]{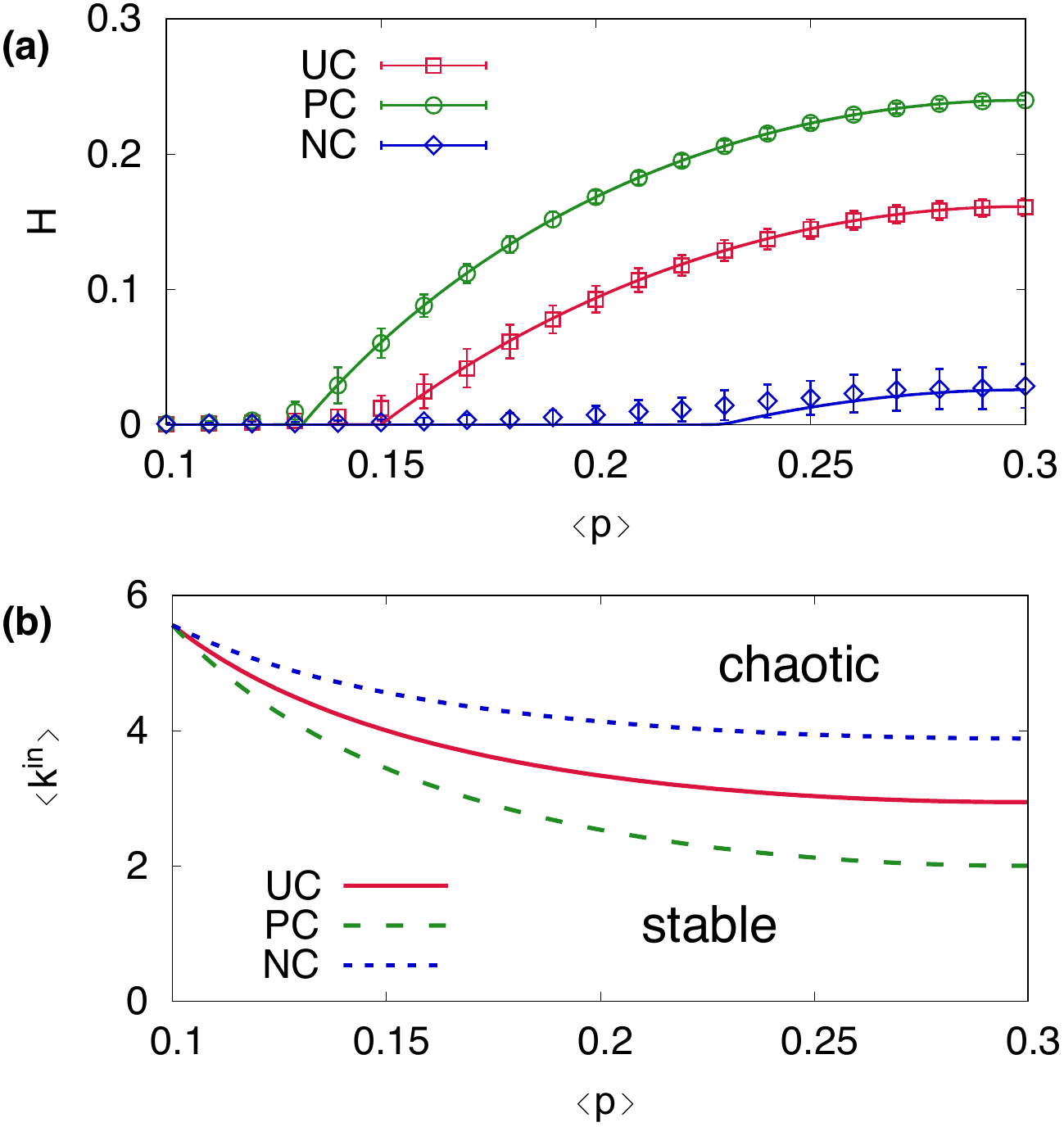}
\caption{
(a) Average Hamming distances of random Boolean networks with 
three types of correlated coupling and a bimodal in-degree distribution 
with $P(k^{in})= (1/2)\delta_{k^{in},10}+(1/2)\delta_{k^{in},6}$ and $N=10^4$.
The bias distribution is also bimodal and defined as 
$Q(p)=(1/2)\delta_{p,2\langle p \rangle - 1/2}+(1/2)\delta_{p,0.1}$.
Analytical predictions (lines) and numerical results (symbols) are 
presented together. 
(b) Phase diagram between stable and chaotic phases obtained
theoretically as a function of the mean node in-degree $\langle k^{in} \rangle$ and $p$. 
We use $K_1=\langle k^{in} \rangle + 2$ and $K_2=\langle k^{in} \rangle -2$.
}
\label{fig:bi}
\end{figure}

\begin{figure*}[t]
\includegraphics[width=\linewidth]{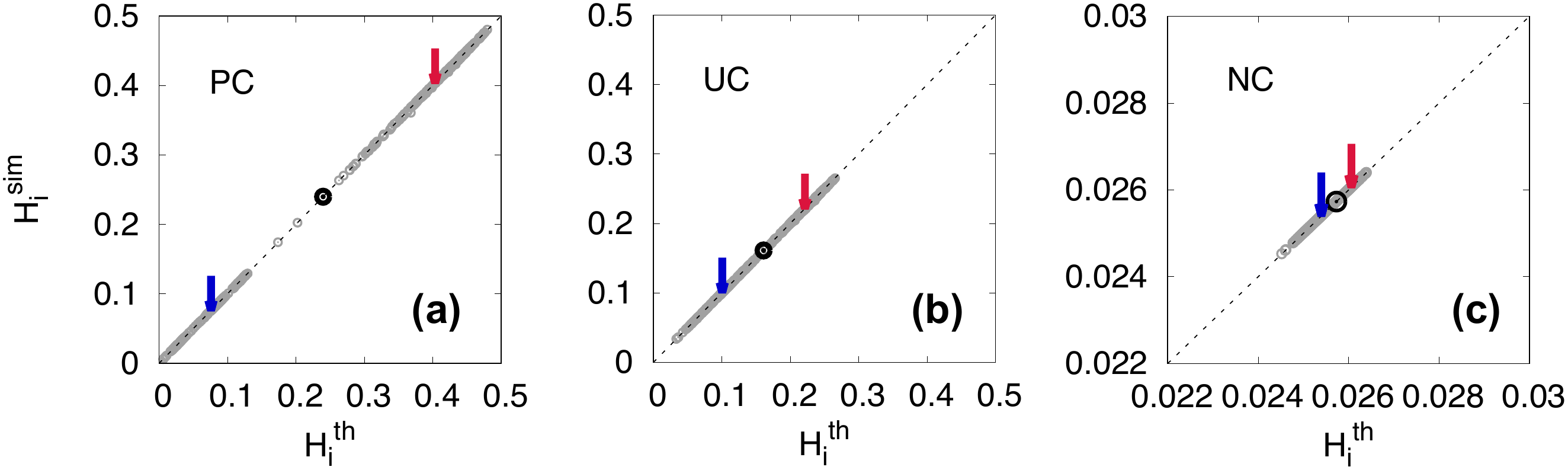}
\caption{
Comparison of the stability of each node obtained 
theoretically $H_i^{th}$ and through numerical simulations $H_i^{sim}$
with different types of coupling: (a) PC, (b) UC, and (c) NC.
We consider a network with $P(k^{in})= (1/2)\delta_{k^{in},10}+(1/2)\delta_{k^{in},6}$ 
and $N=10^4$, and a bimodal bias distribution defined as
$Q(p)=(1/2)\delta_{p,2\langle p \rangle - 1/2}+(1/2)\delta_{p,0.1}$.
Note the changed axes for (c).
The average Hamming distances for nodes with high in-degree ($k^{in}=10$) and low 
degree ($k^{in}=6$) are denoted by red and blue arrows, respectively.
The average Hamming distance over all nodes is denoted by a filled circle. 
}
\label{fig:nodes}
\end{figure*}

To evaluate the impact of the interplay between node degree and sensitivity more 
clearly, we consider a transparent example with a bimodal in-degree distribution 
$P(k^{in})= (1/2)\delta_{k^{in},K_1}+(1/2)\delta_{k^{in},K_2}$, where $\delta_{i,j}$ 
represents the Kronecker delta. We construct networks with the bimodal in-degree 
and Poisson distributed out-degree distribution according to the configuration model.
We assign a bias drawn from a bimodal distribution 
$Q(p)=(1/2)\delta_{p,\phi_1}+(1/2)\delta_{p,\phi_2}$. 
Similar to the analysis above, we consider three types of correlated coupling: UC,
PC, and NC. For the UC case, we assign a bias to each node at random, 
independent of the node degree. Positively (negatively) correlated 
coupling can be achieved by ensuring that higher (lower) degree nodes have 
greater bias values.
In our examples, we use $K_1=10$, $K_2=6$, $\phi_1=2\langle p\rangle-1/2$, 
and $\phi_2=0.1$, where $0.1 \le  \phi_1 \le 0.5$. Note that the range of 
$\langle p \rangle$ is $0.1 \le \langle p \rangle \le 0.3$.
By annealing the probability $H_i$, we can derive the self-consistency equation 
for $H_a$ as follows:
\begin{align}
\text{UC:} \ H_a &= \frac{1}{2} \left( q_1  + q_2 \right) \left[2-(1-H_a)^{K_1}-(1-H_a)^{K_2} \right], \nonumber \\
\text{PC:} \ H_a &= \frac{1}{2} q_1 \left[1-(1-{H_a})^{K_1} \right] + \frac{1}{2} q_2 \left[1-(1-{H_a})^{K_2} \right],  \nonumber \\
\text{NC:} \ H_a &= \frac{1}{2} q_2 \left[1-(1-{H_a})^{K_1} \right] + \frac{1}{2} q_1 \left[1-(1-{H_a})^{K_2} \right], 
\end{align}
where $q_1=2 \phi_1 (1-\phi_1)$ and $q_2=2 \phi_2 (1-\phi_2)$.

As shown in Fig.~\ref{fig:bi}, negatively correlated coupling is more resilient to external 
perturbations compared to the other types of coupling. The average Hamming distance 
clearly highlights the effect of the correlation between sensitivity and local node in-degree. 
Specifically, negative correlation between sensitivity and node degree enhances the 
global stability of Boolean networks [Fig.~\ref{fig:bi}(a)]. 
For the NC case, the majority of incoming links are connected to unbiased nodes ($p=1/2$), 
leading to more stable Boolean dynamics. In contrast, for the PC case, damage can easily 
spread through an entire network because an adequate fraction of high-degree nodes 
have high sensitivity. Fig.~\ref{fig:bi}(b) presents a phase diagram as functions of 
the mean node degree $\langle k^{in}\rangle$ and $\langle p\rangle$, where 
$K_1=\langle k^{in} \rangle + 2$ and $K_2=\langle k^{in} \rangle -2$.
An increasing value of $p_c$ for the NC coupling can be 
observed consistently over a wide range of parameter sets.

In addition to the global stability of Boolean networks, we can also assess
the stability of each node in a given network topology using Eq.~\ref{eq:hi}. 
Fig.~\ref{fig:nodes} reveals perfect agreement between the numerical results 
$H_i^{sim}$ and theoretical predictions $H_i^{th}$ for the probability 
that a node $i$ changes its state when a perturbation occurs. 
We computed the average Hamming distances $\langle H(k^{in}=10) \rangle$ 
and $\langle H(k^{in}=6) \rangle$ for nodes with high in-degree ($k^{in}=10$) and 
low in-degree ($k^{in}=6$), respectively, as indicated by the red and blue arrows 
in Fig.~\ref{fig:nodes}, respectively. The average Hamming distance 
over all nodes is denoted by a filled circle. One can see two clearly separated 
groups of nodes with different stability values and Hamming distances $H_i$ for 
the PC coupling. We can confirm that in the PC coupling, damage can spread through 
high in-degree nodes with high sensitivity, which are prone to instability.
However, for the NC coupling, these two groups merge and perturbations terminate 
quickly. From the perspective of a percolation problem, NC coupling corresponds
to the case where high in-degree nodes have low occupation probabilities,
leading to stable dynamics, which is analogous to degree-based removal 
in network percolation \cite{attack}.

Finally, we consider the role of the interplay between local node in-degree
and canalizing inputs \cite{moreira2005,daniels2018}. 
Canalizing functions have a single input that forces the 
corresponding output to a specific value, regardless of the values 
of other inputs. The Hamming distance $H_i$ for each node $i$ with 
a fraction $c_i$ of canalizing inputs is calculated as \cite{squires2012}
\begin{align}
H_{i} = & c_i q_i H_{ic} + \frac{1}{2} c_i q_i (1-H_{ic}) 
	\left[ 1- \prod_{j \in \partial i/c} (1-H_{j}) \right] \nonumber \\ 
&\quad +(1-c_i) q_i  \left[ 1- \prod_{j \in \partial i} (1-H_{j}) \right], 
\label{eq:canal}
\end{align}
where $H_{ic}$ is the Hamming distance of the neighbor of node $i$
connected by a canalizing input, $q_j=2p_j(1-p_j)$, 
$\partial j/c$ define a set of inputs excluding 
a canalizing input \cite{squires2012}. By definition, canalizing functions 
lead to stable dynamics \cite{moreira2005} because they effectively reduce 
the sensitivity of non-canalizing inputs connected to nodes shared by 
canalizing inputs. However, the effects of canalizing inputs are not solely 
determined by the fraction of canalizing inputs. 
The topological locations of canalizing links also affect stability,
which can be predicted using Eq.~\ref{eq:canal}.

\begin{figure}
\includegraphics[width=\linewidth]{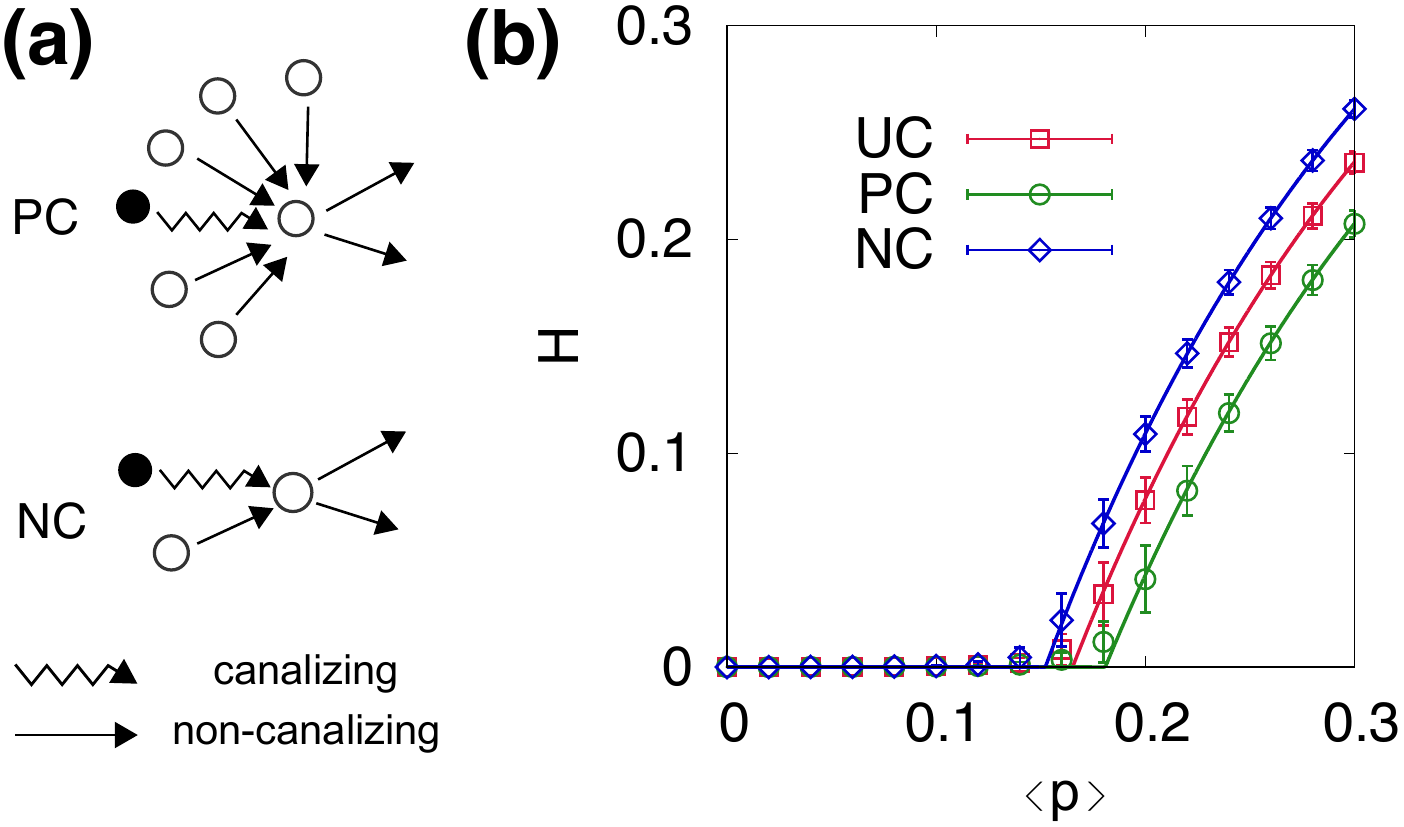}
\caption{
(a) Diagram of coupling between node degree and canalizing 
inputs. For the PC and NC cases, nodes with high and low in-degree 
respectively tend to have a canalizing input.
(b) Average Hamming distance with canalizing inputs 
as a function of $\langle p\rangle$ with the UC, PC, and NC cases.
}
\label{fig:canal}
\end{figure}

We consider three different correlations between node in-degree and the locations 
of canalizing inputs, which are again denoted as UC, PC, and NC. For the UC case, 
the canalizing inputs are distributed randomly. For the PC (NC) case 
nodes with high (low) in-degree tend to have a canalizing input.
For the sake of simplicity, we consider a random network with a bimodal in-degree 
distribution defined as $P(k^{in})= (1/2)\delta_{k^{in},K_1}+(1/2)\delta_{k^{in},K_2}$, 
where $K_1=6$ and $K_2=2$. In this example, we assume that $1/4$ of the nodes 
have a canalizing input and each node can have at most one canalizing input.
For the UC case, a quarter of nodes chosen at random have a canalizing input.
For the PC case, the low in-degree nodes ($k^{in}=2$) do not have a canalizing input
and a half of high in-degree nodes ($k^{in}=6$) have a canalizing input. 
On the other hand, for the NC case only low in-degree nodes have 
a canalizing input with the probability $1/2$ [see Fig.~\ref{fig:canal}(a)].

As shown in Fig.~\ref{fig:canal}(b), correlation between canalizing inputs and local 
node in-degree can increase global stability. Specifically, the PC coupling enhances the 
stability of Boolean networks. In contrast, the NC coupling decreases stability, leading 
to smaller $p_c$ and larger $H$ values. In this example, one can see that correlation 
between local node degree and canalizing inputs alters Boolean dynamics significantly. 
When a canalizing input becomes active, all other connections are ineffective. 
Therefore, for the PC case, a larger fraction of non-canalizing inputs lose their 
influence on Boolean dynamics to the canalizing inputs. In the NC case, the effects 
of canalizing inputs are minimized because they only affect low in-degree nodes.

\section{Discussion}

We analyzed the stability of random Boolean networks incorporating 
dynamic rules, as well as the topological properties of each node. We find that correlation 
between node degree and Boolean functions plays an important 
role in determining Hamming distances and critical point. Specifically, negatively 
correlated sensitivity to the in-degree of each node increases stability. 
We also find that a correlation between high-degree nodes and canalizing inputs 
can increase global stability. Our results reveal that analysis based on the naive 
mean-field approach may fail to predict dynamical consequences in Boolean networks
with intertwined structural and functional properties, which are often observed 
in real-world biological systems. Further study is required to examine the effects 
of more complex features in Boolean networks, such as loops and feedback in network 
topologies \cite{kinoshita,cho}, and hierarchical dynamics.

\section*{Acknowledgments}
This work was supported by the National Research Foundation of
Korea (NRF) grant funded by the Korean government (MSIT) (NRF-2018R1C1B5044202 
and NRF-2020R1I1A3068803).

\section*{Data Availability}
The data that supports the findings of this study are available within the article.


\section*{References}

\begin{thebibliography}{1000}
\bibitem{kauffman1969} S. A. Kauffman, 
	{\it J. Theoret. Biol.} {\bf 22}, 437 (1969). 
\bibitem{karlebach2008} G. Karlebach and R. Shamir, 
	{\it Nat. Rev. Mol. Cell Biol.} {\bf 9}, 770 (2008). 
\bibitem{albert2008} I. Albert, J. Thakar, S. Li, R. Zhang, and R. Albert, 
	{\it Source Code Biol. Med.} {\bf 3}, 16 (2008). 
\bibitem{li2004} F. Li, T. Long, Y. Lu, Q. Ouyang, and C. Tang, 
	{\it Proc. Natl. Acad. Sci.} {\bf 101}, 4781 (2004). 
\bibitem{thomas2001} R. Thomas and M. Kaufman, 
	{\it Chaos} {\bf 11}, 180-195 (2001). 
\bibitem{shmulevich2002} I. Shmulevich, E. R. Dougherty, and W. Zhang, 
	{\it Bioinformatics} {\bf 18}, 1319 (2002). 
\bibitem{rosin2013} D. P. Rosin, D. Rontani, D. J. Gauthier, and E. Sch\"oll, 
	{\it Phys. Rev. Lett.} {\bf 110}, 104102 (2013). 
\bibitem{sayama2013} H. Sayama, I. Pestov, J. Schmidt, B. J. Bush, C. Wong, J. Yamanoi, and T. Gross, 
	{\it Comput. Math. Appl.} {\bf 65}, 1645 (2013).
\bibitem{kauffmanbook} S. A. Kauffman,
	The origins of order: self-organization and selection in evolution,
	(Oxford University Press, New York, 1993).
\bibitem{munoz2018} M. A. M\~unoz, {\it Rev. Mod. Phys.} 
	{\bf 90}, 031001 (2018). 
\bibitem{krotov2014} D. Krotov, J. O. Dubuis, T. gregor, and W. Bialek, 
	{\it Proc. Natl. Acad. Sci.} {\bf 111}, 3683 (2014).
\bibitem{daniels2018} B. C. Daniels, H. Kim, D. Moore, S. Zhou, H. B. Smith, B. Karas,
	S. A. Kauffman, and S. I. Walker, 
	{\it Phys. Rev. Lett.} {\bf 121}, 138102 (2018). 
\bibitem{serra2004} R. Serra, M. Villani, and A. Semeria, 
	{\it J. Theor. Biol.} {\bf 227}, 149 (2004). 
\bibitem{nykter2008} M. Nykter, N. D. Price, M. Aldana, S. A. Ramsey, S. A. Kauffman, 
	L. E. Hood, O. Yli-Harja, and I. Shmulevich, 
	{\it Proc. Natl. Acad. Sci.} {\bf 105}, 1897 (2008). 
\bibitem{derrida1986} B. Derrida and Y. Pomeau, 
	{\it EPL (Europhys. Lett.)} {\bf 1}, 45 (1986). 
\bibitem{luque1997} B. Luque and R. V. Sole, 
	{\it Phys. Rev. E} {\bf 55}, 257 (1997).
\bibitem{aldana2002} M. Aldana and P. Cluzel, 
	{\it Proc. Natl. Acad. Sci.} {\bf 100}, 8710 (2003). 
\bibitem{lee2008} D.-S. Lee and H. Rieger, 
	{\it J. Phys. A: Math. Theor.} {\bf 41}, 415001 (2008). 
\bibitem{peixoto2009} T. P. Peixoto and B. Drossel, 
	{\it Phys. Rev. E} {\bf 79}, 036108 (2009). 
\bibitem{villegas2016} P. Villegas, J. Ruiz-Franco, J. Hidalgo, and M. A. Mu\~noz, 
	{\it Sci. Rep.} {\bf 6}, 34743 (2016). 
\bibitem{cozzo2012} E. Cozzo, A. Arenas, and Y. Moreno, 
	{\it Phys. Rev. E} {\bf 86}, 036115 (2012). 
\bibitem{klemm2005} K. Klemm and S. Bornholdt, 
	{\it Phys. Rev. E} {\bf 72}, 055101(R) (2005). 
\bibitem{ghanbarnejad2011} F. Ghanbarnejad and K. Klemm, 
	{\it Phys. Rev. Lett.} {\bf 107}, 188701 (2011). 
\bibitem{ebadi2014} H. Ebadi and K. Klemm, 
	{\it Phys. Rev. E} {\bf 90}, 022815 (2014). 
\bibitem{lee2012} D. Lee, K.-I. Goh, and B. Kahng, 
	{\it Phys. Rev. E} {\bf 86}, 027101 (2012). 
\bibitem{boldhaus2010} G. Boldhaus, N. Bertschinger, J. Rauh, E. Olbrich, and K. Klemm, 
	{\it Phys. Rev. E} {\bf 82}, 021916 (2010). 
\bibitem{wang2018} J. Wang, S. Pei, W. Wei, X. Feng, and Z. Zheng, 
	{\it Phys. Rev. E} {\bf 97}, 032305 (2018). 
\bibitem{goudarzi2012} A. Goudarzi, C. Teuscher, N. Gulbahce, and T. Rohlf, 
	{\it Phys. Rev. Lett.} {\bf 108}, 128702 (2012). 
\bibitem{moreira2005} A. A. Moreira and L. A. N. Amaral, 
	{\it Phys. Rev. Lett.} {\bf 94}, 218702 (2005). 
\bibitem{shmulevich2004} I. Shmulevich and S. A. Kauffman, 
	{\it Phys. Rev. Lett.} {\bf 93(4)}, 048701 (2004). 
\bibitem{squires2014} S. Squires, A. Pomerance, M. Girvan, and E. Ott, 
	{\it Phys. Rev. E} {\bf 90}, 022814 (2014). 
\bibitem{pomerance2009} A. Pomerance, E. Ott, M. Girvan, and W. Losert, 
	{\it Proc. Natl. Acad. Sci.} {\bf 106}, 8209 (2009). 
\bibitem{squires2012} S. Squires, E. Ott, and M. Girvan,
	{\it Phys. Rev. Lett.} {\bf 108}, 085701 (2012). 
\bibitem{bmin} B. Min,
	{\it Eur. Phys. J. B} {\bf 91}, 18 (2018). 
\bibitem{bmin2} B. Min and C. Castellano,
	{\it Chaos} {\bf 30}, 023131 (2020). 
\bibitem{conroy} B. D. Conroy, et. al., 
	{\it Front. Immunol.} {\bf 5}, 599 (2014).
\bibitem{guelzim} N. Guelzim, S. Bottani, P. Bourgine, and F. Kepes,
	{\it Nat. Genet.} {\bf 31}, 60 (2002).
\bibitem{dobrin} R. Dobrin, Q. K. Beg, A.-L. Barab\'asi, and Z. N. Oltvai,
	{\it BMC Bioinformatics} {\bf 5}, 10 (2004).
\bibitem{assortative} M. E. J. Newman,
	{\it Phys. Rev. Lett.} {\bf 89}, 208701 (2002).
\bibitem{interlayer} B. Min, S. D. Yi, K.-M. Lee, and K.-I. Goh,
	{\it Phys. Rev. E} {\bf 89}, 042811 (2014).
\bibitem{attack} R. Albert, H. Jeong, and A.-L. Barab\'asi, 
	{\it Nature} {\bf 406}, 378-382 (2000).
\bibitem{kinoshita} S. Kinoshita and H. S. Yamada,
	{\it Open Journal of Biophysics} {\bf 9}, 10-20 (2019).
\bibitem{cho} Y.-K. Kwon and K.-H. Cho,
	{\it BMC Bioinformatics} {\bf 8}, 430 (2007).

\end{thebibliography}
\end{document}